\documentclass[aps,prd,twocolumn,nofootinbib,floatfix,superscriptaddress,longbibliography=false]{revtex4-2}

\usepackage[utf8]{inputenc}
\usepackage{amsmath,amssymb}
\usepackage{graphicx}
\usepackage{hyperref}
\usepackage{physics}
\usepackage{booktabs}
\usepackage{xcolor}
\usepackage{svg}
\DeclareMathAlphabet\mathbfcal{OMS}{cmsy}{b}{n}

\begin{document}

\title{CMB Bounds on Primordial Black Holes via Radiation Capture: Linear Perturbations}

\author{M. Farhang}
%\email{m\_farhang@sbu.ac.ir}
\affiliation{Department of Physics, Shahid Beheshti University, Tehran 1983969411, Iran}

\author{S. M. S. Movahed}
\affiliation{Department of Physics, Shahid Beheshti University, Tehran 1983969411, Iran}
\affiliation{School of Astronomy, Institute for Research in Fundamental Sciences (IPM), P. O. Box, Tehran 19395-5531, Iran}
\affiliation{Department of Mathematics and Statistics, The University of Lahore, 1-KM Defence Road, Lahore 54000, Pakistan}
%\email{m.s.movahed@ipm.ir}

\date{\today}

\begin{abstract}
We investigate the gravitational interaction between primordial radiation, consisting of photons and ultra-relativistic neutrinos, and primordial black holes (PBHs), and study its imprint on the anisotropies of the cosmic microwave background (CMB). Assuming that PBHs are immersed in the cosmic radiation background, we introduce a covariant energy-momentum transfer four-vector to describe the gravitational capture of radiation by PBHs. We extend the background-level framework of \cite{Farhang:2026auc} to an inhomogeneous Universe at linear order by consistently modifying the perturbed continuity and Euler equations to account for both energy and momentum transfer. We use the imprint of this interaction on the CMB to place novel constraints on the PBH abundance, $f_{\rm pbh}$.
We show that these constraints are up to two orders of magnitude tighter than those obtained from background-only analyses and exhibit a strong dependence on the PBH mass. Using {\it Planck} temperature and polarization data, we constrain the PBH fraction to
$f_{\rm pbh} \lesssim 0.05-0.002$ for $M$ in the range $10^{15}-10^{17} M_\odot$.
Furthermore, sensitivity forecasts for a cosmic-variance-limited survey with $\ell_{\rm max}=7000$, assuming a fiducial $f_{\rm pbh}=0$, yield projected bounds of $f_{\rm pbh} \lesssim 10^{-2}-2\times 10^{-4}$ over the mass range $10^{13}-10^{17} M_\odot$.

\end{abstract}
\maketitle

\section{Introduction}\label{sec:intro}
The possibility of black hole formation in the early Universe was first proposed by \cite{Zeldovich:1967lct,Hawking:1971ei}. It  was soon realized that these primordial black holes, or PBHs, could constitute at least a fraction  of dark matter \citep{Chapline:1975ojl,Meszaros:1975ef}. 
PBH masses can span many orders of magnitude, and if existing, their  rich   phenomenology  produces diverse astrophysical and cosmological signatures. These  signatures are used to constrain PBH abundance, often parametrized by $f_{\rm pbh}$,   and include constraints from evaporation \citep{Carr:2009jm}, lensing \citep{Wilkinson:2001vv,Niikura:2017zjd,Niikura:2019kqi,Mediavilla:2009um}, accretion \citep{Ali-Haimoud:2016mbv,Horowitz:2016lib,Poulin:2017bwe,Serpico:2020ehh}, dynamical effects \citep{Montero-Camacho:2019jte,Graham:2015apa,Capela:2013yf,Roncadelli:2009qj}, $\mu$-distortions in the cosmic microwave background (CMB) \citep{Kohri:2014lza} and gravitational waves \citep{Vaskonen:2019jpv,Raidal:2018bbj}. 

In a recent work, \cite{Farhang:2026auc} studied the imprints of gravitational capture of radiation by PBHs on CMB. They considered PBH population embedded in a sea of radiation, and explored how the radiation capture by PBHs would impact the power spectra of CMB temperature and polarization anisotropies. 

The analysis was performed at the background level, treating PBHs and radiation as homogeneous fluids that exchange energy through an interaction term constructed to ensure the energy conservation. This was achieved by modifying the continuity equations for the PBH and radiation fluids. These modifications, in turn, affected background quantities that depended on the radiation and dark matter densities, particularly the Hubble parameter.
 They found that the imprints are only significant for extremely massive black holes, with $M \gtrsim 10^{13} M_\odot $ for a  future cosmic variance limited CMB experiment with $\ell_{\rm max}=7000$. 

In this work we extend their analysis  to the linear perturbation level.
An inhomogeneous treatment requires perturbing the energy-momentum conservation equations, thereby modifying the continuity and Euler equations to account for both energy and momentum transfer between the different species involved. We consider a model where PBHs fully absorb the momenta of the captured radiation (section~\ref{sec:momentumtransfer}).
In contrast, the homogeneous analysis does not involve momentum transfer.
The modified equations would affect the observational imprints of the interactions and lead to different constraints on PBH abundance, as measured or predicted by CMB surveys.

The paper is organized as follows. In Section~\ref{sec:formalism} we introduce the covariant formalism used to model the radiation capture by PBHs, both in the background (Section~\ref{sub:bkgnd}) and perturbed levels (Section~\ref{sub:pert}). We explore the observational consequences of perturbation treatment on CMB anisotropies in Section~\ref{sec:analysis}. The results, i.e., the constraint on PBH abundance for various PBH masses, are presented and discussed in Section~\ref{sec:res}. We conclude and finish with a brief discussion in Section~\ref{sec:discussion}.

\section{Covariant Interaction formalism}\label{sec:formalism}
In this section we introduce our proposed model for the gravitational capture of radiation, i.e., photons and ultra-relativistic  neutrinos, by PBHs in the early Universe. We effectively model the radiation and PBH components as  fluids, and  describe their interaction through an energy-momentum transfer four-vector. 
Throughout this work we assume a monochromatic PBH distribution. 
%\subsubsection{gauge and notation}

In this work we choose to work in synchronous gauge where scalar perturbations to metric are characterized as
\begin{equation}
ds^2=a^2[-d\tau^2+(\delta_{ij}+h_{ij}) dx^idx^j]
\end{equation}
with 
\begin{equation}
h_{ij}=-2\psi \delta_{ij}-2\hat{k}_i\hat{k}_jk^2E
\end{equation}
in Fourier space.
The energy-momentum tensor for fluid A is described by
\begin{equation}
T^\mu_{{\rm A}\nu}=(\rho_{\rm A}+P_{\rm A})u^\mu_{\rm A} u_\nu^{\rm A}+P_{\rm A}\delta^\mu_\nu+\pi^\mu_{{\rm A}\nu}
\end{equation}
where $\rho_{\rm A}$,  $P_{\rm A}$ and $\pi^\mu_{{\rm A}\nu}$ represent the density, pressure and anisotropic stress of fluid A. We take $\pi^\mu_{{\rm A}\nu}$ to be zero for PBHs and photons in the tightly coupled regime, but is non-negligible for free-streaming neutrinos.
The total energy--momentum tensor is conserved,
\begin{equation}
\nabla_\mu T^{\mu\nu}=0, ~~ T^{\mu\nu}=  \sum_{\rm A} T^{\mu\nu}_{\rm A}
\end{equation}
with ${\rm A=pbh, g,n}$ labelling PBHs, photons and neutrinos, respectively.
We describe the energy and momentum exchange due to radiation capture  by the interaction four-vector $Q_{\rm A}^\mu$, satisfying
\begin{align}\label{eq:conservation}
\nabla_\mu T^{\mu\nu}_{\rm A} &=Q_{\rm A}^\nu
\end{align}
with $Q_{\rm pbh}^\mu=-(Q_{\rm g}^\mu+Q_{\rm n}^\mu)$.
Each interaction vector can be decomposed into energy and momentum transfer components,
\begin{equation}\label{eq:Qgeneral}
Q_{\rm A}^\mu=Q_{\rm A} u^\mu+F_{\rm A}^\mu,
\end{equation}
where
\begin{equation}
u_\mu F_{\rm A}^\mu=0.
\end{equation}
Here $u^\mu$ denotes the total four-velocity,
\begin{equation}\label{eq:vtotal}
u^\mu =a^{-1} \left( 1,\, \partial^i v \right),
\end{equation}
where $v$ is the peculiar velocity potential, defined through
\begin{equation}
(\rho+P)v =\sum_{\rm A} (\rho_{\rm A}+P_{\rm A})v_{\rm A}
\end{equation}
and  $\rho=\sum_{\rm A}\rho_{\rm A}$ and $P=\sum_{\rm A} P_{\rm A}$.
The momentum transfer potential $\zeta_{\rm A}$ for component A is introduced through
\begin{equation}\label{eq:Fmu}
F_{\rm A}^\mu
=a^{-1}\left(0,\,\partial^i \zeta_{\rm A}\right).
\end{equation}
Our choice for momentum transfer would be fully determined by  $\zeta_{\rm A}$  (see sections~\ref{sec:momentumtransfer}). 
Conservation of the energy-momentum tensor imposes $\zeta_{\rm pbh}=-(\zeta_{\rm g}+\zeta_{\rm n})$.

In \cite{Farhang:2026auc} the authors studied the fluid description of radiation capture by PBHs in the homogeneous limit, and explored whether its impact on CMB anisotropies could constrain PBH abundance. In the following, we first re-derive the background evolution equations for PBH and radiation densities in the presence of the interaction term (Section~\ref{sub:bkgnd}) and then extend the analysis to the linear regime of perturbations (Section~\ref{sub:pert}).

\iffalse
The 4-vector $Q^\mu$ describes the interaction between radiation and pbhs, with 
\begin{eqnarray}
\nabla_\nu T^{\mu \nu}_{\rm pbh}&=&Q^\mu_{\rm n}+Q^\mu_{\rm g}=Q^\mu_{\rm pbh}\\ \nonumber 
\nabla_\nu T^{\mu \nu}_{\rm x}&=&-Q^\mu_{\rm x}; ~~ {\rm x=g,n}
\end{eqnarray}
where
\begin{equation}\label{eq:Q1}
Q^\mu_{\rm x}=Q_{\rm x} u_\mu + F^\mu_{\rm x}; ~~~ u^\mu F^\mu=0
\end{equation}
where $u^\mu$ is the total four-velocity
\begin{equation}\label{eq:4veltot}
u^\mu=a^{-1}(1,\partial^i v).
\end{equation}
Here $v$ is defined through
\begin{equation}
(\rho+P)v=\sum_{\rm A} \left( \rho_{\rm A}+P_{\rm A}\right)  v_{\rm A}.
\end{equation}
%
We also define the momentum transfer potential as $F_{\rm x}^\mu=a^{-1}(0,\partial^i f_{\rm x})$ 
\fi
%----------------------------
\subsection{Background equations}\label{sub:bkgnd}
In the homogeneous background level, PBHs are embedded in an isotropic radiation sea and the net momentum transfer is zero. Therefore, PBHs and radiation only exchange energy.
By projecting the conservation equation along $u_\mu$  and keeping the homogeneous part (represented by overbar) we get
\begin{equation}
\dot{{\bar \rho}}_{\rm A} +3H(1+w_{\rm A}){\bar \rho}_{\rm A}= {\bar Q}_{\rm A}. 
\end{equation}
where 
\begin{eqnarray}
H(z)&=&H_0 \Big[ \Omega_{\rm b}(1+z)^3 + (1-f_{\rm pbh})\Omega_{\rm dm}(1+z)^3\nonumber  \\
&+& \Omega_{\rm pbh}(z)+ \Omega_{\rm g} (z) + \Omega_{\rm n} (z) + \Omega_\Lambda \Big]^{1/2},
\end{eqnarray}
with $\Omega_{{\rm b, dm}, \Lambda}$ labelling the current density parameters associated to the baryons, dark matter and cosmological constant, and  $\Omega_{\rm A}(z)$ representing the density parameters of neutrinos, photons and PBhs  at redshift $z$. 
Comparing with Eq. 3 of \cite{Farhang:2026auc}, we set 
\begin{equation}
{\bar Q}_{\rm x}=-C A_{\rm x} {\bar \rho}_{\rm x} {\bar\rho}_{\rm pbh}^2, ~~ {\rm x= g, n}
\end{equation}
and ${\bar Q}_{\rm pbh}=-({\bar Q}_{\rm n}+{\bar Q}_{\rm g} )$ 
where $C=\frac{\pi G^2}{\rho_{\rm dm, 0}(1+z)^3} \frac{M}{f_{\rm pbh}}$,  $M$ is PBH mass,  $\rho_{\rm dm, 0}$ is the current dark matter energy density,  and $A_{\rm g}\approx 45$ and $A_{\rm n}=27$ \citep{Farhang:2026auc}. 

%----------------------------------
\subsection{Linear perturbations} \label{sub:pert}
%----------------------------------
In this section we extend the analysis of PBH-radiation gravitational interaction to the inhomogeneous Universe in the linear regime.  
Here, beside the perturbed continuity equation, perturbed Euler equation also needs to be considered.

%----------------------------------
\subsubsection{Continuity and Euler equations}
%----------------------------------
Consider fluids with sound speed $c_{\rm s}^2=w$ where the equation of state $w$ is constant.
Perturbing continuity equation in Fourier space for such fluids yields 
\begin{equation}\label{eq:continuitygeneral}
\delta'_{\rm A}+(1+w_{\rm A})\theta_{\rm A}+\frac{1}{2}(1+w_{\rm A})h'
=\frac{a}{{\bar \rho}_{\rm A}}\left[\delta Q_{\rm A}-{\bar Q}_{\rm A} \delta_{\rm A}  \right]
\end{equation}
where $h=-6\psi -2k^2E$, $\theta=-k^2 v$ and primes represent derivatives with respect to conformal time $\tau$ (see Eq.~30 of \cite{Valiviita:2008iv}, and set $\phi=B=0$ for synchronous gauge).
We have $c_{\rm s, n}^2=w_{\rm n}=1/3$,  and $c_{\rm s, pbh}^2=w_{\rm pbh}=0$. For photons we have $w_{\rm g}=1/3$ and for the sound speed of the coupled baryon-photon fluid we have $c_{\rm s, g}^2=w_{\rm g}=1/(3+4R)$ where $R$ is the baryon-to-photon ratio. The interaction is expected to be most significant deep in the radiation era, where R is tiny so we safely set $c_{\rm s, g}^2=1/3$.

For the perturbed interaction rate we have 
\begin{eqnarray}
\delta Q_{\rm x} &=&\bar{Q}_{\rm x} (\delta_{\rm x}+2\delta_{\rm pbh}), ~~~ {\rm x=g,n}\nonumber \\
\delta Q_{\rm pbh}&=&-(\delta Q_{\rm g}+\delta Q_{\rm n}).
\end{eqnarray}
Therefore  the perturbed continuity equation, Eq. ~\ref{eq:continuitygeneral}, for PBHs and radiation components reduces to
\begin{eqnarray}
\delta'_{\rm pbh}+\theta_{\rm pbh}+\frac{h'}{2}
&=&  \frac{a}{\rho_{\rm pbh}}(\delta_{\rm pbh}{\bar Q}_{\rm pbh}- \sum\bar{Q}_{\rm x}\delta_{\rm x})  \label{eq:continpbh} \\ 
\delta'_{\rm x}+\frac{4}{3}\theta_{\rm x}+\frac{2}{3}h'
&=&\frac{2a}{\bar{\rho}_{\rm x}}  {\bar Q}_{\rm x}\delta_{\rm pbh}, ~~~ {\rm x = g, n.} \label{eq:continx} 
\end{eqnarray}
%In the following, we formulate our model for the momentum transfer, as needed in the Euler equation (Eq.~\ref{eq:eulergeneral}).
%
For Euler equations we have (see Eq.~31 of \cite{Valiviita:2008iv})
\begin{eqnarray} \label{eq:eulergeneral}
&&\theta'_{\rm A}+{\cal H}(1-3c_{\rm s,A}^2)\theta_{\rm A}-\frac{c_{\rm s,A}^2}{1+w_{\rm A}}k^2\delta_{\rm A}+\frac{2}{3a^2(1+w_{\rm A}){\bar \rho}_{\rm A}}k^4\pi_{\rm A} \nonumber \\
&&= \left[ \frac{a{\bar Q}_{\rm A}}{(1+w_{\rm A}){\bar\rho}_{\rm A}}\left[\theta-(1+c_{{\rm s},{\rm A}}^2)\theta_{\rm A}\right] - \frac{a}{(1+w_{\rm A})\rho_{\rm A}}k^2 \zeta_{\rm A}  \right]   \nonumber 
\end{eqnarray}
for A=pbh, g and n, and with ${\cal H}=aH$.
Eq.~\ref{eq:eulergeneral} is the general perturbed Euler equation for interacting fluids. To proceed, we should specify the momentum transfer potential $\zeta_{\rm A}$, which is not fixed by background dynamics and only enters at the level of perturbations. We adopt the physically-motivated assumption that PBHs fully absorb the momentum of the captured radiation, as detailed below.

We note that the interaction, introduced phenomenologically through the covariant energy-momentum transfer four-vector $Q^\mu$, only modifies the continuity and Euler equations within the interacting-fluid formalism.  The evolution equations for higher moments of the neutrino distribution, including the anisotropic stress, are assumed to remain identical to those of collisionless neutrinos.
This can also be verified at the level of Boltzmann equations with a proper choice of collision term, corresponding to the assumed  $Q^\mu$.

%------------------------
\subsubsection{Momentum transfer}\label{sec:momentumtransfer}
The goal in this section is to construct a physically-motivated model for the momentum transfer potential $\zeta_{\rm A}$  which would enter the source term (RHS) of the Euler equations for PBHs and radiation (Eq.~\ref{eq:eulergeneral}). 
For this purpose, first consider the energy-momentum flux of radiation (type ${\rm x}$) as measured by a PBH with four-velocity $u^\mu_{\rm pbh}$, 
\begin{equation}
J^\mu_{\rm x}=-T^{\mu \nu}_{\rm x}u_\nu^{\rm pbh}.
\end{equation}
 We make the physical assumption that  PBHs fully absorb the momenta of all radiation passing through the effective interaction cross section. This would suggest 
\begin{equation}\label{eq:model2}
Q^\mu_{\rm pbh}= -\kappa J^\mu_{\rm pbh}=-\kappa(J_{\rm g}^\mu+J_{\rm n}^\mu), ~~Q_{\rm x}^\mu
=\kappa J_{\rm x}^\mu 
\end{equation}
 The scalar $\kappa$ is determined by matching  $Q^0$ to the  energy transfer rate.
 First note that one can write
$J^\mu_{\rm x}=\rho_{\rm x} u^{\mu}+q_{\rm x}^\mu$ where $\rho_{\rm x}=-u_\mu J_{\rm x}^\mu$  is  the energy density and $q^\mu_{\rm x}=h^\mu_{~\nu} J_{\rm x}^\nu$ is  the  flux of radiation x as measured by an observer with four-velocity $u^\mu$  and $h^\mu_{~\nu}=\delta^\mu_\nu+u^\mu u_\nu$.
 Contracting $Q^\mu_{\rm x}$ as defined in Eq.~\ref{eq:model2} with $-u_{\mu}$ and comparing the result with  Eq.~\ref{eq:Qgeneral} yields $Q_{\rm x}=\kappa (-J_{\rm x}^\mu  u_\mu)=\kappa \rho_{\rm x}$.
%This is equal to the energy density of fluid x in its rest frame (Eq.~\ref{eq:Tmunu}) to $O(v^2)$.
%
Therefore
\begin{equation}
\kappa=\frac{Q_{\rm x}}{\rho_{\rm x}}.
\end{equation}
For $F^\mu_{\rm x}$ we contract both sides of Eq.~\ref{eq:model2} with $h^\mu_{~\nu}$,
\begin{equation}\label{eq:FJ}
F^\mu_{\rm x}=h^\mu_{~\nu} Q_{\rm x}^\nu=h^\mu_{~\nu} (\kappa J_{\rm x}^\nu).
\end{equation}
%
%\frac{Q_{\rm x}}{\rho_{\rm x}} q_{\rm x}^\mu .
Noting that $q_{\rm x}^\mu=h^\mu_{~\nu}  J_{\rm x}^\nu$, we get
\begin{equation}
q^\nu_{\rm x}=a^{-1}\left( 0,\rho_{\rm x}\partial^i( v_{\rm x}-v) +P_{\rm x}\partial^i(v_{\rm x}-v_{\rm pbh})\right).
\end{equation}
Comparing with Eq.~\ref{eq:FJ} and using Eq.~\ref{eq:Fmu} we get
\begin{eqnarray}\label{eq:fx}
\zeta_{\rm x}&=&\frac{Q_{\rm x}}{\rho_{\rm x}}\left[ \rho_{\rm x}(v_{\rm x}-v)+P_{\rm x} (v_{\rm x} -v_{\rm pbh}) \right] \nonumber \\
  &=&Q_{\rm x}\left [ (1+w_{\rm x})v_{\rm x} -v - w_{\rm x} v_{\rm pbh} \right].
\end{eqnarray} 
Then, plugging Eq~\ref{eq:fx} in the RHS of EQ~\ref{eq:eulergeneral} we get for PBHs
\begin{equation}\label{eq:eulerpbh2}
\theta'_{\rm pbh}+{\cal H}\theta_{\rm pbh}=\frac{2a}{{\bar \rho}_{\rm pbh}}\left[\bar{Q}_{\rm pbh}(\theta-\frac{1}{3}\theta_{\rm pbh})+\frac{2}{3}(\bar{Q}_{\rm n}\theta_{\rm n}+\bar{Q}_{\rm g}\theta_{\rm g})\right].
\end{equation}
For massless neutrinos Eq.~\ref{eq:eulergeneral} becomes
\begin{equation}\label{eq:eulerx2}
\theta'_{\rm n}-\frac{1}{4}k^2\delta_{\rm n}+\frac{1}{2a^2 {\bar \rho}_{\rm n}}k^4 \pi_{\rm n }=
\frac{3aQ_{\rm n}}{4\rho_{\rm n}}\left[ 2\theta -\frac{8}{3} \theta_{\rm n} +\frac{1}{3}\theta_{\rm pbh} \right].
\end{equation}
 The same  modification would apply to the source in the  RHS of the Euler equation for photons.
 The modified continuity and Euler equations for PBHs and radiation (Eqs.~\ref{eq:continpbh},  \ref{eq:continx}, ~\ref{eq:eulerpbh2} and \ref{eq:eulerx2}) are the main results of this section. 
%------------------------------
\begin{figure}[t]
	\centering 
		\includegraphics[width=0.45\textwidth]{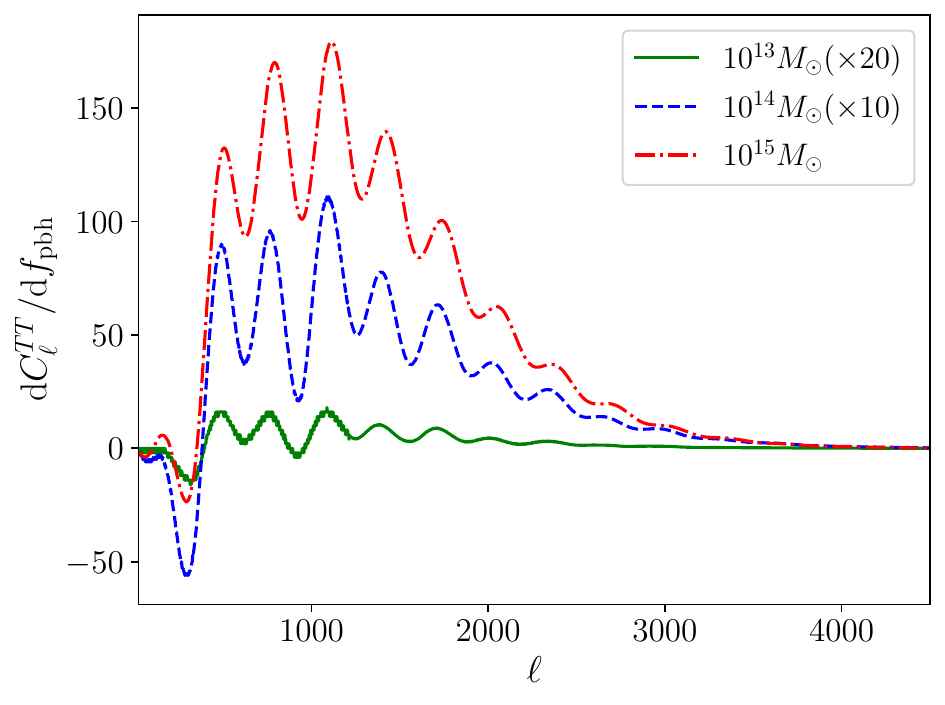}
		\includegraphics[width=0.45\textwidth]{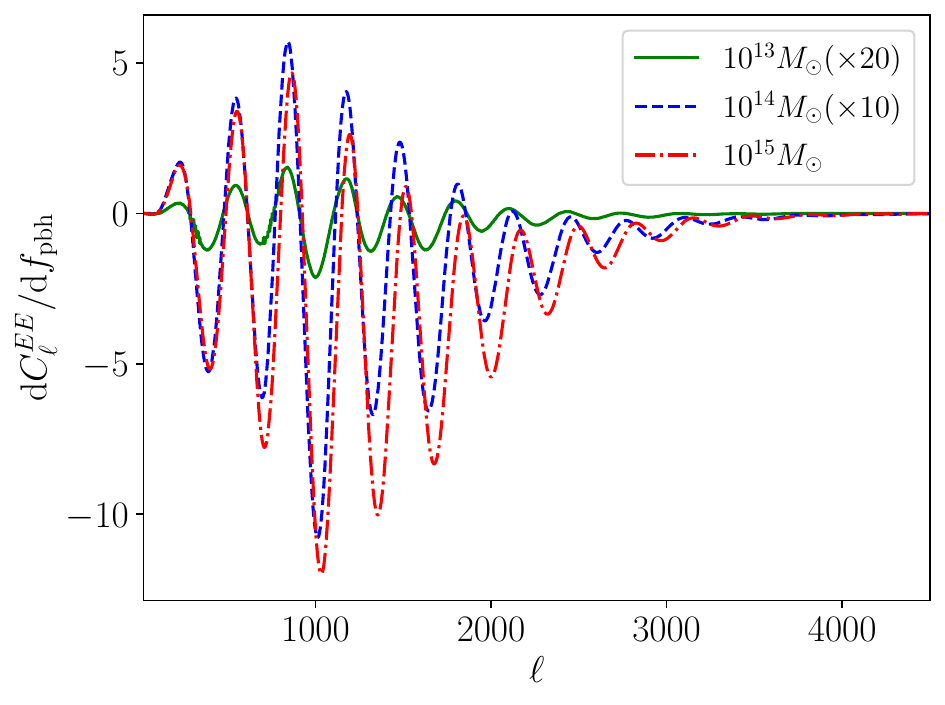}
	\caption{The response of  CMB temperature and $E$-mode polarization power spectra to variations in PBH abundance, with the fiducial $f_{\rm pbh}=0$, and for various PBH mass.}
	\label{fig:dCldp}%
\end{figure}
%-----------------------------
%------------------------------
\begin{figure}[t]
	\centering 
		\includegraphics[width=0.45\textwidth]{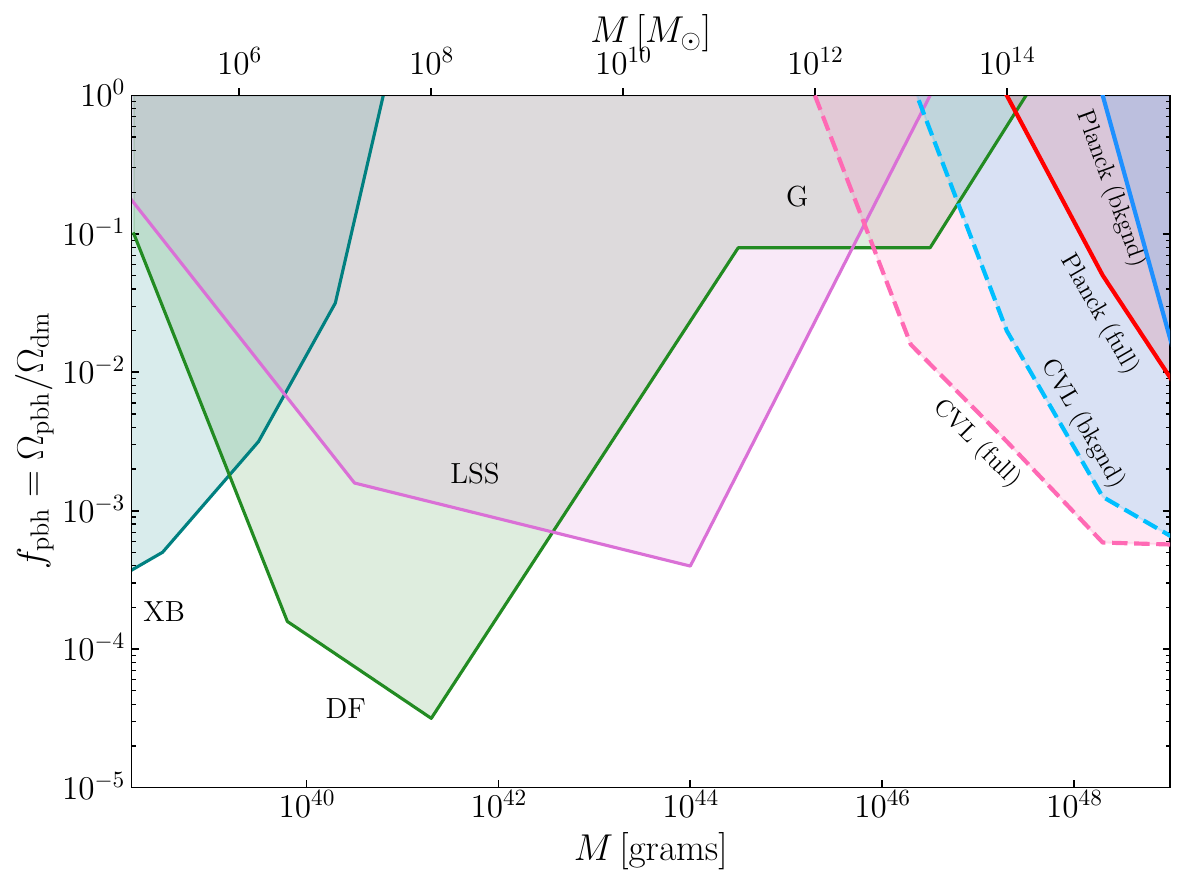}
	\caption{$3\sigma$ upper bounds on PBH abundance from $Planck$ data (solid dark pink), and a cosmic-variance-limited (CVL) CMB experiment with $\ell_{\rm max}=7000$ (dashed light pink), imprinted by PBH-radiation interaction. The analysis extends the background results of \cite{Farhang:2026auc} (in blue) to the level of linear perturbations. Constraints from other surveys are included for comparison (adopted from \cite{Carr:2020erq}), partially based on \cite{Carr:2020xqk}.}
	\label{fig:bound}%
\end{figure}
%-----------------------------
\section{Analysis}\label{sec:analysis}
In this work we use the modified conservation equation of the energy-momentum tensor, in the homogeneous Universe as well as in the linearly perturbed level to take into account the PBH-radiation interaction, as described in Section~\ref{sec:formalism}. We modify the corresponding equations in the 
Boltzmann solver code, CAMB\footnote{https://camb.info/. The modified code is available upon request.}.
The expected sensitivity of the CMB temperature and $E$-mode anisotropies to PBH abundance for various PBH masses are plotted in Figure~\ref{fig:dCldp}. The interaction affects the evolution of radiation and matter densities, and therefore the quantities depending on them, in particular the expansion history of the Universe. Also, the source terms for density and velocity perturbations are coupled and impact the evolution of matter overdensities and radiation anisotropies in nontrivial ways.
It is evident from the figure that  the response is strongly mass-dependent and grows substantially with PBH mass. 

The modified equations are then implemented in the cosmological MCMC code, CosmoMC\footnote{https://cosmologist.info/cosmomc/} (version 2021), to constrain PBH abundance, $f_{\rm pbh}$, which is added as a free parameter to the standard set of cosmological parameters. We use anisotropies of CMB temperature and polarization for two observational cases: {\it Planck} (plik\_rd12\_HM\_v22\_TTTEEE, lowl and simall\_EE likelihood packages), and a cosmic variance limited CMB experiment (CVL) up to $\ell_{\rm max}=7000$.
We assume flat prior on $\log f_{\rm pbh}$ in the range $[-20,0]$.
As the  convergence criterion for CosmoMC runs we use the Gelman-Rubin statistic with $R-1 \lesssim 0.01$.
The main results are the $3\sigma$ upper bounds on $f_{\rm pbh}$, marginalized over standard cosmological parameters,  and are  presented in Section~\ref{sec:res}.

%----------------------------
\section{Results}\label{sec:res}
Our goal in this work is to use the cosmological imprints of radiation-PBH interaction as discussed in the previous sections to constrain PBH abundance. Figure~\ref{fig:bound} illustrates the $3\sigma$ upper bound on $f_{\rm pbh}$, imposed by {\it Planck} data (red solid line) and as predicted for a CVL experiment with fiducial $f_{\rm pbh}=0$ (red dashed line), for various PBH mass scales. 
 The bounds are tightest for larger PBH masses, and only non-trivial for $M \gtrsim 10^{14}M_\odot$ for {\it Planck} and  $M \gtrsim 10^{12}M_\odot$ for CVL case. 
The {\it Planck}
 sets $f_{\rm pbh} \lesssim 0.05-0.002$ for $M$ in $10^{15}-10^{17} M_\odot$. 
With a CVL experiment with $\ell_{\rm max}=7000$ we expect the minimum detectable trace to be from PBHs with $M \sim 10^{12} M_\odot$, and we get $f_{\rm pbh} \lesssim 10^{-2}-2\times 10^{-4}$ for $M$ in $10^{13}-10^{17} M_\odot$.
These bounds, depending on the assumed PBH mass, are up to two orders of magnitude tighter  than the bounds found from the background-only level of PBH-radiation interaction scenario \citep{Farhang:2026auc}. This is expected, as incorporating the modification to the perturbed continuity and Euler equations would add to the imprints left by the interaction on cosmological observables. 
We also find negligible correlation between $f_{\rm pbh}$ and standard cosmological parameters.

\section{Conclusion and discussion} \label{sec:discussion}
In this work we investigate the imprints of radiation capture by PBHs on the CMB temperature and polarization power spectra. We utilize these signatures to derive  upper limits on the PBH abundance. This work  extends the analysis of \cite{Farhang:2026auc}, where the required modifications were only studied in the homogeneous limit, to the level of linear perturbations. The bounds on $f_{\rm pbh}$ are found to be considerably tighter than the homogeneous-only analysis. However, the PBHs constrained through the imprint of their gravitational interaction with radiation are extremely massive. Such objects are often referred to as stupendously large black holes, or SLABs. A primordial origin for black holes of this mass remains controversial, as discussed in \cite{Farhang:2026auc}. See also \cite{Carr:2020erq} for an extensive review of SLABs and their astrophysical consequences. We emphasize, however, that the primary goal of this work is not to constrain the abundance of SLABs, but rather to extend the homogeneous treatment of PBH--radiation gravitational interactions to the more realistic regime of linear cosmological perturbations.

It should also be noted that the validity of the fluid approximation is less clear for PBHs at the high-mass end of the spectrum, where the discrete nature of PBHs  becomes non-negligible. 
Also, for such objects, and immediately after their formation, the  steady-state approximation for radiation capture is least reliable. See \cite{Farhang:2026auc} for more discussion. We leave a fully time-dependent treatment of the capture process to future work.

\bibliography{refs}

\begin{thebibliography}{24}%
\makeatletter
\providecommand \@ifxundefined [1]{%
 \@ifx{#1\undefined}
}%
\providecommand \@ifnum [1]{%
 \ifnum #1\expandafter \@firstoftwo
 \else \expandafter \@secondoftwo
 \fi
}%
\providecommand \@ifx [1]{%
 \ifx #1\expandafter \@firstoftwo
 \else \expandafter \@secondoftwo
 \fi
}%
\providecommand \natexlab [1]{#1}%
\providecommand \enquote  [1]{``#1''}%
\providecommand \bibnamefont  [1]{#1}%
\providecommand \bibfnamefont [1]{#1}%
\providecommand \citenamefont [1]{#1}%
\providecommand \href@noop [0]{\@secondoftwo}%
\providecommand \href [0]{\begingroup \@sanitize@url \@href}%
\providecommand \@href[1]{\@@startlink{#1}\@@href}%
\providecommand \@@href[1]{\endgroup#1\@@endlink}%
\providecommand \@sanitize@url [0]{\catcode `\\12\catcode `\$12\catcode
  `\&12\catcode `\#12\catcode `\^12\catcode `\_12\catcode `\%12\relax}%
\providecommand \@@startlink[1]{}%
\providecommand \@@endlink[0]{}%
\providecommand \url  [0]{\begingroup\@sanitize@url \@url }%
\providecommand \@url [1]{\endgroup\@href {#1}{\urlprefix }}%
\providecommand \urlprefix  [0]{URL }%
\providecommand \Eprint [0]{\href }%
\providecommand \doibase [0]{https://doi.org/}%
\providecommand \selectlanguage [0]{\@gobble}%
\providecommand \bibinfo  [0]{\@secondoftwo}%
\providecommand \bibfield  [0]{\@secondoftwo}%
\providecommand \translation [1]{[#1]}%
\providecommand \BibitemOpen [0]{}%
\providecommand \bibitemStop [0]{}%
\providecommand \bibitemNoStop [0]{.\EOS\space}%
\providecommand \EOS [0]{\spacefactor3000\relax}%
\providecommand \BibitemShut  [1]{\csname bibitem#1\endcsname}%
\let\auto@bib@innerbib\@empty
%</preamble>
\bibitem [{\citenamefont {Farhang}\ and\ \citenamefont
  {Movahed}(2026)}]{Farhang:2026auc}%
  \BibitemOpen
  \bibfield  {author} {\bibinfo {author} {\bibfnamefont {M.}~\bibnamefont
  {Farhang}}\ and\ \bibinfo {author} {\bibfnamefont {S.~M.~S.}\ \bibnamefont
  {Movahed}},\ }\bibfield  {title} {\bibinfo {title} {{CMB Bounds on Primordial
  Black Holes via Radiation Capture}},\ }\href@noop {} {\bibfield  {journal}
  {\bibinfo  {journal} {Phys. Rev. D}\ }\textbf {\bibinfo {volume} {114}},\
  \bibinfo {pages} {083533} (\bibinfo {year} {2026})},\ \Eprint
  {https://arxiv.org/abs/2606.05133} {arXiv:2606.05133 [astro-ph.CO]}
  \BibitemShut {NoStop}%
\bibitem [{\citenamefont {Zel'dovich}\ and\ \citenamefont
  {Novikov}(1967)}]{Zeldovich:1967lct}%
  \BibitemOpen
  \bibfield  {author} {\bibinfo {author} {\bibfnamefont {Y.~B.}\ \bibnamefont
  {Zel'dovich}}\ and\ \bibinfo {author} {\bibfnamefont {I.~D.}\ \bibnamefont
  {Novikov}},\ }\bibfield  {title} {\bibinfo {title} {{The Hypothesis of Cores
  Retarded during Expansion and the Hot Cosmological Model}},\ }\href@noop {}
  {\bibfield  {journal} {\bibinfo  {journal} {Sov. Astron.}\ }\textbf {\bibinfo
  {volume} {10}},\ \bibinfo {pages} {602} (\bibinfo {year} {1967})}\BibitemShut
  {NoStop}%
\bibitem [{\citenamefont {Hawking}(1971)}]{Hawking:1971ei}%
  \BibitemOpen
  \bibfield  {author} {\bibinfo {author} {\bibfnamefont {S.}~\bibnamefont
  {Hawking}},\ }\bibfield  {title} {\bibinfo {title} {{Gravitationally
  collapsed objects of very low mass}},\ }\href
  {https://doi.org/10.1093/mnras/152.1.75} {\bibfield  {journal} {\bibinfo
  {journal} {Mon. Not. Roy. Astron. Soc.}\ }\textbf {\bibinfo {volume} {152}},\
  \bibinfo {pages} {75} (\bibinfo {year} {1971})}\BibitemShut {NoStop}%
\bibitem [{\citenamefont {Chapline}(1975)}]{Chapline:1975ojl}%
  \BibitemOpen
  \bibfield  {author} {\bibinfo {author} {\bibfnamefont {G.~F.}\ \bibnamefont
  {Chapline}},\ }\bibfield  {title} {\bibinfo {title} {{Cosmological effects of
  primordial black holes}},\ }\href {https://doi.org/10.1038/253251a0}
  {\bibfield  {journal} {\bibinfo  {journal} {Nature}\ }\textbf {\bibinfo
  {volume} {253}},\ \bibinfo {pages} {251} (\bibinfo {year}
  {1975})}\BibitemShut {NoStop}%
\bibitem [{\citenamefont {Meszaros}(1975)}]{Meszaros:1975ef}%
  \BibitemOpen
  \bibfield  {author} {\bibinfo {author} {\bibfnamefont {P.}~\bibnamefont
  {Meszaros}},\ }\bibfield  {title} {\bibinfo {title} {{Primeval black holes
  and galaxy formation}},\ }\href@noop {} {\bibfield  {journal} {\bibinfo
  {journal} {Astron. Astrophys.}\ }\textbf {\bibinfo {volume} {38}},\ \bibinfo
  {pages} {5} (\bibinfo {year} {1975})}\BibitemShut {NoStop}%
\bibitem [{\citenamefont {Carr}\ \emph {et~al.}(2010)\citenamefont {Carr},
  \citenamefont {Kohri}, \citenamefont {Sendouda},\ and\ \citenamefont
  {Yokoyama}}]{Carr:2009jm}%
  \BibitemOpen
  \bibfield  {author} {\bibinfo {author} {\bibfnamefont {B.~J.}\ \bibnamefont
  {Carr}}, \bibinfo {author} {\bibfnamefont {K.}~\bibnamefont {Kohri}},
  \bibinfo {author} {\bibfnamefont {Y.}~\bibnamefont {Sendouda}},\ and\
  \bibinfo {author} {\bibfnamefont {J.}~\bibnamefont {Yokoyama}},\ }\bibfield
  {title} {\bibinfo {title} {{New cosmological constraints on primordial black
  holes}},\ }\href {https://doi.org/10.1103/PhysRevD.81.104019} {\bibfield
  {journal} {\bibinfo  {journal} {Phys. Rev. D}\ }\textbf {\bibinfo {volume}
  {81}},\ \bibinfo {pages} {104019} (\bibinfo {year} {2010})},\ \Eprint
  {https://arxiv.org/abs/0912.5297} {arXiv:0912.5297 [astro-ph.CO]}
  \BibitemShut {NoStop}%
\bibitem [{\citenamefont {Wilkinson}\ \emph {et~al.}(2001)\citenamefont
  {Wilkinson}, \citenamefont {Henstock}, \citenamefont {Browne}, \citenamefont
  {Polatidis}, \citenamefont {Augusto}, \citenamefont {Readhead}, \citenamefont
  {Pearson}, \citenamefont {Xu}, \citenamefont {Taylor},\ and\ \citenamefont
  {Vermeulen}}]{Wilkinson:2001vv}%
  \BibitemOpen
  \bibfield  {author} {\bibinfo {author} {\bibfnamefont {P.~N.}\ \bibnamefont
  {Wilkinson}}, \bibinfo {author} {\bibfnamefont {D.~R.}\ \bibnamefont
  {Henstock}}, \bibinfo {author} {\bibfnamefont {I.~W.~A.}\ \bibnamefont
  {Browne}}, \bibinfo {author} {\bibfnamefont {A.~G.}\ \bibnamefont
  {Polatidis}}, \bibinfo {author} {\bibfnamefont {P.}~\bibnamefont {Augusto}},
  \bibinfo {author} {\bibfnamefont {A.~C.~S.}\ \bibnamefont {Readhead}},
  \bibinfo {author} {\bibfnamefont {T.~J.}\ \bibnamefont {Pearson}}, \bibinfo
  {author} {\bibfnamefont {W.}~\bibnamefont {Xu}}, \bibinfo {author}
  {\bibfnamefont {G.~B.}\ \bibnamefont {Taylor}},\ and\ \bibinfo {author}
  {\bibfnamefont {R.~C.}\ \bibnamefont {Vermeulen}},\ }\bibfield  {title}
  {\bibinfo {title} {{Limits on the cosmological abundance of supermassive
  compact objects from a search for multiple imaging in compact radio
  sources}},\ }\href {https://doi.org/10.1103/PhysRevLett.86.584} {\bibfield
  {journal} {\bibinfo  {journal} {Phys. Rev. Lett.}\ }\textbf {\bibinfo
  {volume} {86}},\ \bibinfo {pages} {584} (\bibinfo {year} {2001})},\ \Eprint
  {https://arxiv.org/abs/astro-ph/0101328} {arXiv:astro-ph/0101328}
  \BibitemShut {NoStop}%
\bibitem [{\citenamefont {Niikura}\ \emph
  {et~al.}(2019{\natexlab{a}})\citenamefont {Niikura} \emph
  {et~al.}}]{Niikura:2017zjd}%
  \BibitemOpen
  \bibfield  {author} {\bibinfo {author} {\bibfnamefont {H.}~\bibnamefont
  {Niikura}} \emph {et~al.},\ }\bibfield  {title} {\bibinfo {title}
  {{Microlensing constraints on primordial black holes with Subaru/HSC
  Andromeda observations}},\ }\href {https://doi.org/10.1038/s41550-019-0723-1}
  {\bibfield  {journal} {\bibinfo  {journal} {Nature Astron.}\ }\textbf
  {\bibinfo {volume} {3}},\ \bibinfo {pages} {524} (\bibinfo {year}
  {2019}{\natexlab{a}})},\ \Eprint {https://arxiv.org/abs/1701.02151}
  {arXiv:1701.02151 [astro-ph.CO]} \BibitemShut {NoStop}%
\bibitem [{\citenamefont {Niikura}\ \emph
  {et~al.}(2019{\natexlab{b}})\citenamefont {Niikura}, \citenamefont {Takada},
  \citenamefont {Yokoyama}, \citenamefont {Sumi},\ and\ \citenamefont
  {Masaki}}]{Niikura:2019kqi}%
  \BibitemOpen
  \bibfield  {author} {\bibinfo {author} {\bibfnamefont {H.}~\bibnamefont
  {Niikura}}, \bibinfo {author} {\bibfnamefont {M.}~\bibnamefont {Takada}},
  \bibinfo {author} {\bibfnamefont {S.}~\bibnamefont {Yokoyama}}, \bibinfo
  {author} {\bibfnamefont {T.}~\bibnamefont {Sumi}},\ and\ \bibinfo {author}
  {\bibfnamefont {S.}~\bibnamefont {Masaki}},\ }\bibfield  {title} {\bibinfo
  {title} {{Constraints on Earth-mass primordial black holes from OGLE 5-year
  microlensing events}},\ }\href {https://doi.org/10.1103/PhysRevD.99.083503}
  {\bibfield  {journal} {\bibinfo  {journal} {Phys. Rev. D}\ }\textbf {\bibinfo
  {volume} {99}},\ \bibinfo {pages} {083503} (\bibinfo {year}
  {2019}{\natexlab{b}})},\ \Eprint {https://arxiv.org/abs/1901.07120}
  {arXiv:1901.07120 [astro-ph.CO]} \BibitemShut {NoStop}%
\bibitem [{\citenamefont {Mediavilla}\ \emph {et~al.}(2009)\citenamefont
  {Mediavilla}, \citenamefont {Munoz}, \citenamefont {Falco}, \citenamefont
  {Motta}, \citenamefont {Guerras}, \citenamefont {Canovas}, \citenamefont
  {Jean}, \citenamefont {Oscoz},\ and\ \citenamefont
  {Mosquera}}]{Mediavilla:2009um}%
  \BibitemOpen
  \bibfield  {author} {\bibinfo {author} {\bibfnamefont {E.}~\bibnamefont
  {Mediavilla}}, \bibinfo {author} {\bibfnamefont {J.~A.}\ \bibnamefont
  {Munoz}}, \bibinfo {author} {\bibfnamefont {E.}~\bibnamefont {Falco}},
  \bibinfo {author} {\bibfnamefont {V.}~\bibnamefont {Motta}}, \bibinfo
  {author} {\bibfnamefont {E.}~\bibnamefont {Guerras}}, \bibinfo {author}
  {\bibfnamefont {H.}~\bibnamefont {Canovas}}, \bibinfo {author} {\bibfnamefont
  {C.}~\bibnamefont {Jean}}, \bibinfo {author} {\bibfnamefont {A.}~\bibnamefont
  {Oscoz}},\ and\ \bibinfo {author} {\bibfnamefont {A.~M.}\ \bibnamefont
  {Mosquera}},\ }\bibfield  {title} {\bibinfo {title} {{Microlensing-Based
  Estimate of the Mass Fraction in Compact Objects in Lens}},\ }\href
  {https://doi.org/10.1088/0004-637X/706/2/1451} {\bibfield  {journal}
  {\bibinfo  {journal} {Astrophys. J.}\ }\textbf {\bibinfo {volume} {706}},\
  \bibinfo {pages} {1451} (\bibinfo {year} {2009})},\ \Eprint
  {https://arxiv.org/abs/0910.3645} {arXiv:0910.3645 [astro-ph.CO]}
  \BibitemShut {NoStop}%
\bibitem [{\citenamefont {Ali-Ha{\"\i}moud}\ and\ \citenamefont
  {Kamionkowski}(2017)}]{Ali-Haimoud:2016mbv}%
  \BibitemOpen
  \bibfield  {author} {\bibinfo {author} {\bibfnamefont {Y.}~\bibnamefont
  {Ali-Ha{\"\i}moud}}\ and\ \bibinfo {author} {\bibfnamefont {M.}~\bibnamefont
  {Kamionkowski}},\ }\bibfield  {title} {\bibinfo {title} {{Cosmic microwave
  background limits on accreting primordial black holes}},\ }\href
  {https://doi.org/10.1103/PhysRevD.95.043534} {\bibfield  {journal} {\bibinfo
  {journal} {Phys. Rev. D}\ }\textbf {\bibinfo {volume} {95}},\ \bibinfo
  {pages} {043534} (\bibinfo {year} {2017})},\ \Eprint
  {https://arxiv.org/abs/1612.05644} {arXiv:1612.05644 [astro-ph.CO]}
  \BibitemShut {NoStop}%
\bibitem [{\citenamefont {Horowitz}(2016)}]{Horowitz:2016lib}%
  \BibitemOpen
  \bibfield  {author} {\bibinfo {author} {\bibfnamefont {B.}~\bibnamefont
  {Horowitz}},\ }\bibfield  {title} {\bibinfo {title} {{Revisiting Primordial
  Black Holes Constraints from Ionization History}},\ }\href@noop {} {\
  (\bibinfo {year} {2016})},\ \Eprint {https://arxiv.org/abs/1612.07264}
  {arXiv:1612.07264 [astro-ph.CO]} \BibitemShut {NoStop}%
\bibitem [{\citenamefont {Poulin}\ \emph {et~al.}(2017)\citenamefont {Poulin},
  \citenamefont {Serpico}, \citenamefont {Calore}, \citenamefont {Clesse},\
  and\ \citenamefont {Kohri}}]{Poulin:2017bwe}%
  \BibitemOpen
  \bibfield  {author} {\bibinfo {author} {\bibfnamefont {V.}~\bibnamefont
  {Poulin}}, \bibinfo {author} {\bibfnamefont {P.~D.}\ \bibnamefont {Serpico}},
  \bibinfo {author} {\bibfnamefont {F.}~\bibnamefont {Calore}}, \bibinfo
  {author} {\bibfnamefont {S.}~\bibnamefont {Clesse}},\ and\ \bibinfo {author}
  {\bibfnamefont {K.}~\bibnamefont {Kohri}},\ }\bibfield  {title} {\bibinfo
  {title} {{CMB bounds on disk-accreting massive primordial black holes}},\
  }\href {https://doi.org/10.1103/PhysRevD.96.083524} {\bibfield  {journal}
  {\bibinfo  {journal} {Phys. Rev. D}\ }\textbf {\bibinfo {volume} {96}},\
  \bibinfo {pages} {083524} (\bibinfo {year} {2017})},\ \Eprint
  {https://arxiv.org/abs/1707.04206} {arXiv:1707.04206 [astro-ph.CO]}
  \BibitemShut {NoStop}%
\bibitem [{\citenamefont {Serpico}\ \emph {et~al.}(2020)\citenamefont
  {Serpico}, \citenamefont {Poulin}, \citenamefont {Inman},\ and\ \citenamefont
  {Kohri}}]{Serpico:2020ehh}%
  \BibitemOpen
  \bibfield  {author} {\bibinfo {author} {\bibfnamefont {P.~D.}\ \bibnamefont
  {Serpico}}, \bibinfo {author} {\bibfnamefont {V.}~\bibnamefont {Poulin}},
  \bibinfo {author} {\bibfnamefont {D.}~\bibnamefont {Inman}},\ and\ \bibinfo
  {author} {\bibfnamefont {K.}~\bibnamefont {Kohri}},\ }\bibfield  {title}
  {\bibinfo {title} {{Cosmic microwave background bounds on primordial black
  holes including dark matter halo accretion}},\ }\href
  {https://doi.org/10.1103/PhysRevResearch.2.023204} {\bibfield  {journal}
  {\bibinfo  {journal} {Phys. Rev. Res.}\ }\textbf {\bibinfo {volume} {2}},\
  \bibinfo {pages} {023204} (\bibinfo {year} {2020})},\ \Eprint
  {https://arxiv.org/abs/2002.10771} {arXiv:2002.10771 [astro-ph.CO]}
  \BibitemShut {NoStop}%
\bibitem [{\citenamefont {Montero-Camacho}\ \emph {et~al.}(2019)\citenamefont
  {Montero-Camacho}, \citenamefont {Fang}, \citenamefont {Vasquez},
  \citenamefont {Silva},\ and\ \citenamefont
  {Hirata}}]{Montero-Camacho:2019jte}%
  \BibitemOpen
  \bibfield  {author} {\bibinfo {author} {\bibfnamefont {P.}~\bibnamefont
  {Montero-Camacho}}, \bibinfo {author} {\bibfnamefont {X.}~\bibnamefont
  {Fang}}, \bibinfo {author} {\bibfnamefont {G.}~\bibnamefont {Vasquez}},
  \bibinfo {author} {\bibfnamefont {M.}~\bibnamefont {Silva}},\ and\ \bibinfo
  {author} {\bibfnamefont {C.~M.}\ \bibnamefont {Hirata}},\ }\bibfield  {title}
  {\bibinfo {title} {{Revisiting constraints on asteroid-mass primordial black
  holes as dark matter candidates}},\ }\href
  {https://doi.org/10.1088/1475-7516/2019/08/031} {\bibfield  {journal}
  {\bibinfo  {journal} {JCAP}\ }\textbf {\bibinfo {volume} {08}},\ \bibinfo
  {pages} {031}},\ \Eprint {https://arxiv.org/abs/1906.05950} {arXiv:1906.05950
  [astro-ph.CO]} \BibitemShut {NoStop}%
\bibitem [{\citenamefont {Graham}\ \emph {et~al.}(2015)\citenamefont {Graham},
  \citenamefont {Rajendran},\ and\ \citenamefont {Varela}}]{Graham:2015apa}%
  \BibitemOpen
  \bibfield  {author} {\bibinfo {author} {\bibfnamefont {P.~W.}\ \bibnamefont
  {Graham}}, \bibinfo {author} {\bibfnamefont {S.}~\bibnamefont {Rajendran}},\
  and\ \bibinfo {author} {\bibfnamefont {J.}~\bibnamefont {Varela}},\
  }\bibfield  {title} {\bibinfo {title} {{Dark Matter Triggers of
  Supernovae}},\ }\href {https://doi.org/10.1103/PhysRevD.92.063007} {\bibfield
   {journal} {\bibinfo  {journal} {Phys. Rev. D}\ }\textbf {\bibinfo {volume}
  {92}},\ \bibinfo {pages} {063007} (\bibinfo {year} {2015})},\ \Eprint
  {https://arxiv.org/abs/1505.04444} {arXiv:1505.04444 [hep-ph]} \BibitemShut
  {NoStop}%
\bibitem [{\citenamefont {Capela}\ \emph {et~al.}(2013)\citenamefont {Capela},
  \citenamefont {Pshirkov},\ and\ \citenamefont {Tinyakov}}]{Capela:2013yf}%
  \BibitemOpen
  \bibfield  {author} {\bibinfo {author} {\bibfnamefont {F.}~\bibnamefont
  {Capela}}, \bibinfo {author} {\bibfnamefont {M.}~\bibnamefont {Pshirkov}},\
  and\ \bibinfo {author} {\bibfnamefont {P.}~\bibnamefont {Tinyakov}},\
  }\bibfield  {title} {\bibinfo {title} {{Constraints on primordial black holes
  as dark matter candidates from capture by neutron stars}},\ }\href
  {https://doi.org/10.1103/PhysRevD.87.123524} {\bibfield  {journal} {\bibinfo
  {journal} {Phys. Rev. D}\ }\textbf {\bibinfo {volume} {87}},\ \bibinfo
  {pages} {123524} (\bibinfo {year} {2013})},\ \Eprint
  {https://arxiv.org/abs/1301.4984} {arXiv:1301.4984 [astro-ph.CO]}
  \BibitemShut {NoStop}%
\bibitem [{\citenamefont {Roncadelli}\ \emph {et~al.}(2009)\citenamefont
  {Roncadelli}, \citenamefont {Treves},\ and\ \citenamefont
  {Turolla}}]{Roncadelli:2009qj}%
  \BibitemOpen
  \bibfield  {author} {\bibinfo {author} {\bibfnamefont {M.}~\bibnamefont
  {Roncadelli}}, \bibinfo {author} {\bibfnamefont {A.}~\bibnamefont {Treves}},\
  and\ \bibinfo {author} {\bibfnamefont {R.}~\bibnamefont {Turolla}},\
  }\bibfield  {title} {\bibinfo {title} {{Primordial black holes are again on
  the limelight}},\ }\href@noop {} {\  (\bibinfo {year} {2009})},\ \Eprint
  {https://arxiv.org/abs/0901.1093} {arXiv:0901.1093 [astro-ph.CO]}
  \BibitemShut {NoStop}%
\bibitem [{\citenamefont {Kohri}\ \emph {et~al.}(2014)\citenamefont {Kohri},
  \citenamefont {Nakama},\ and\ \citenamefont {Suyama}}]{Kohri:2014lza}%
  \BibitemOpen
  \bibfield  {author} {\bibinfo {author} {\bibfnamefont {K.}~\bibnamefont
  {Kohri}}, \bibinfo {author} {\bibfnamefont {T.}~\bibnamefont {Nakama}},\ and\
  \bibinfo {author} {\bibfnamefont {T.}~\bibnamefont {Suyama}},\ }\bibfield
  {title} {\bibinfo {title} {{Testing scenarios of primordial black holes being
  the seeds of supermassive black holes by ultracompact minihalos and CMB
  $\mu$-distortions}},\ }\href {https://doi.org/10.1103/PhysRevD.90.083514}
  {\bibfield  {journal} {\bibinfo  {journal} {Phys. Rev. D}\ }\textbf {\bibinfo
  {volume} {90}},\ \bibinfo {pages} {083514} (\bibinfo {year} {2014})},\
  \Eprint {https://arxiv.org/abs/1405.5999} {arXiv:1405.5999 [astro-ph.CO]}
  \BibitemShut {NoStop}%
\bibitem [{\citenamefont {Vaskonen}\ and\ \citenamefont
  {Veerm{\"a}e}(2020)}]{Vaskonen:2019jpv}%
  \BibitemOpen
  \bibfield  {author} {\bibinfo {author} {\bibfnamefont {V.}~\bibnamefont
  {Vaskonen}}\ and\ \bibinfo {author} {\bibfnamefont {H.}~\bibnamefont
  {Veerm{\"a}e}},\ }\bibfield  {title} {\bibinfo {title} {{Lower bound on the
  primordial black hole merger rate}},\ }\href
  {https://doi.org/10.1103/PhysRevD.101.043015} {\bibfield  {journal} {\bibinfo
   {journal} {Phys. Rev. D}\ }\textbf {\bibinfo {volume} {101}},\ \bibinfo
  {pages} {043015} (\bibinfo {year} {2020})},\ \Eprint
  {https://arxiv.org/abs/1908.09752} {arXiv:1908.09752 [astro-ph.CO]}
  \BibitemShut {NoStop}%
\bibitem [{\citenamefont {Raidal}\ \emph {et~al.}(2019)\citenamefont {Raidal},
  \citenamefont {Spethmann}, \citenamefont {Vaskonen},\ and\ \citenamefont
  {Veerm{\"a}e}}]{Raidal:2018bbj}%
  \BibitemOpen
  \bibfield  {author} {\bibinfo {author} {\bibfnamefont {M.}~\bibnamefont
  {Raidal}}, \bibinfo {author} {\bibfnamefont {C.}~\bibnamefont {Spethmann}},
  \bibinfo {author} {\bibfnamefont {V.}~\bibnamefont {Vaskonen}},\ and\
  \bibinfo {author} {\bibfnamefont {H.}~\bibnamefont {Veerm{\"a}e}},\
  }\bibfield  {title} {\bibinfo {title} {{Formation and Evolution of Primordial
  Black Hole Binaries in the Early Universe}},\ }\href
  {https://doi.org/10.1088/1475-7516/2019/02/018} {\bibfield  {journal}
  {\bibinfo  {journal} {JCAP}\ }\textbf {\bibinfo {volume} {02}},\ \bibinfo
  {pages} {018}},\ \Eprint {https://arxiv.org/abs/1812.01930} {arXiv:1812.01930
  [astro-ph.CO]} \BibitemShut {NoStop}%
\bibitem [{\citenamefont {Valiviita}\ \emph {et~al.}(2008)\citenamefont
  {Valiviita}, \citenamefont {Majerotto},\ and\ \citenamefont
  {Maartens}}]{Valiviita:2008iv}%
  \BibitemOpen
  \bibfield  {author} {\bibinfo {author} {\bibfnamefont {J.}~\bibnamefont
  {Valiviita}}, \bibinfo {author} {\bibfnamefont {E.}~\bibnamefont
  {Majerotto}},\ and\ \bibinfo {author} {\bibfnamefont {R.}~\bibnamefont
  {Maartens}},\ }\bibfield  {title} {\bibinfo {title} {{Instability in
  interacting dark energy and dark matter fluids}},\ }\href
  {https://doi.org/10.1088/1475-7516/2008/07/020} {\bibfield  {journal}
  {\bibinfo  {journal} {JCAP}\ }\textbf {\bibinfo {volume} {07}},\ \bibinfo
  {pages} {020}},\ \Eprint {https://arxiv.org/abs/0804.0232} {arXiv:0804.0232
  [astro-ph]} \BibitemShut {NoStop}%
\bibitem [{\citenamefont {Carr}\ \emph {et~al.}(2021)\citenamefont {Carr},
  \citenamefont {Kuhnel},\ and\ \citenamefont {Visinelli}}]{Carr:2020erq}%
  \BibitemOpen
  \bibfield  {author} {\bibinfo {author} {\bibfnamefont {B.}~\bibnamefont
  {Carr}}, \bibinfo {author} {\bibfnamefont {F.}~\bibnamefont {Kuhnel}},\ and\
  \bibinfo {author} {\bibfnamefont {L.}~\bibnamefont {Visinelli}},\ }\bibfield
  {title} {\bibinfo {title} {{Constraints on Stupendously Large Black Holes}},\
  }\href {https://doi.org/10.1093/mnras/staa3651} {\bibfield  {journal}
  {\bibinfo  {journal} {Mon. Not. Roy. Astron. Soc.}\ }\textbf {\bibinfo
  {volume} {501}},\ \bibinfo {pages} {2029} (\bibinfo {year} {2021})},\ \Eprint
  {https://arxiv.org/abs/2008.08077} {arXiv:2008.08077 [astro-ph.CO]}
  \BibitemShut {NoStop}%
\bibitem [{\citenamefont {Carr}\ and\ \citenamefont
  {Kuhnel}(2020)}]{Carr:2020xqk}%
  \BibitemOpen
  \bibfield  {author} {\bibinfo {author} {\bibfnamefont {B.}~\bibnamefont
  {Carr}}\ and\ \bibinfo {author} {\bibfnamefont {F.}~\bibnamefont {Kuhnel}},\
  }\bibfield  {title} {\bibinfo {title} {{Primordial Black Holes as Dark
  Matter: Recent Developments}},\ }\href
  {https://doi.org/10.1146/annurev-nucl-050520-125911} {\bibfield  {journal}
  {\bibinfo  {journal} {Ann. Rev. Nucl. Part. Sci.}\ }\textbf {\bibinfo
  {volume} {70}},\ \bibinfo {pages} {355} (\bibinfo {year} {2020})},\ \Eprint
  {https://arxiv.org/abs/2006.02838} {arXiv:2006.02838 [astro-ph.CO]}
  \BibitemShut {NoStop}%
\end{thebibliography}%
\end{document}